\documentclass[11pt]{article}
\usepackage{graphicx}
\usepackage{amssymb}
\usepackage{enumerate}

\begin{document}

\title{Aggregation Architecture for Data Reduction and Privacy in Advanced Metering Infrastructure}
\author{James~Christopher~Foreman and Franklin~Pacheco\\Purdue University 401 N. Grant St. West Lafayette, IN 47907}
\date{10 Nov 2015}

\maketitle

\section*{Abstract}
\textbf{Advanced Metering Infrastructure (AMI) have rapidly become a topic of international interest as governments have sponsored their deployment for the purposes of utility service reliability and efficiency, e.g., water and electricity conservation. Two problems plague such deployments. First is the protection of consumer privacy. Second is the problem of huge amounts of data from such deployments. A new architecture is proposed to address these problems through the use of Aggregators, which incorporate temporary data buffering and the modularization of utility grid analysis. These Aggregators are used to deliver anonymized summary data to the central utility while preserving billing and automated connection services. }

\section{Introduction}
Advanced Metering Infrastructure (AMI) comprises an automated collection of IT networked metering devices, i.e., smart meters, used for the measurement of several utility service variables and typically located at the point of service, e.g., at the customer. AMI is most often associated with smart power meters, but AMI also includes water, natural gas, and other metered utility services delivered in a geo-physically distributed network. As a result of their utility service application, AMI networks may comprise hundreds of thousands to millions of distributed smart meters within a given service area, resulting in a massive network. Since each smart meter reports several data points at a frequency on the order of minutes, a large quantity of data is also involved.

\subsection{The Data Flood Problem}
In the past several years, there has been an effort by regulators to strongly incentivize utilities to deploy AMI \cite{joskow2012creating}. Once AMI is launched, up to millions of data points begin to flood the utility with little preparation for how to handle this data. This data flood problem results in a data set that is too cumbersome for effective analysis. While big data tools provide a means by which this data can be analyzed, there is still too much data redundancy through which to wade. In this case, a centralized model of data collection and analysis begins to fail.

Others \cite{lohrmann2011processing,rusitschka2010smart} have looked to the cloud to solve the data flood problem. In these cases, the investigations focus on technologies to accommodate the data flood rather than to reduce it in the field. Distributed processing may remain a capability, but this does not alleviate the troublesome communications and archiving requirements. Also, privacy remains a concern, even further exacerbated through the use of a required cloud server, which may or may not be third party.
What is needed is a distributed approach such that much meaningful analysis can be done in the field and aggregated. The utility need only archive this reduced data set, permitting analysis to be performed much more efficiently.

\subsection{The Privacy Problem}
Another problem arises in AMI, that of privacy. The collection and analysis of individual utility service data can easily reveal much personal information about the customer being served \cite{molina2010private}. This is an international problem and threatens the acceptance of AMI technology by utility customers, i.e., the general population \cite{schoechle2012getting,ehrhardt2010advanced}. Many benefits \cite{ehrhardt2010advanced,depuru2011smart} exist for the deployment of AMI, such as energy efficiency, reliable delivery, and reduced costs, yet these are lost if customers reject \cite{depuru2011smart,evans2011opt} AMI.

Some work has been attempted to solve the privacy problem. The use of aggregation is discussed in \cite{kursawe2011privacy}, but without distributed processing necessary to maintain the fidelity of grid operating state estimation. In \cite{rajagopalan2011smart}, a framework employing a stationary Gaussian Markov model of load and filtering frequency components that are low in power is investigated. While this may provide some distortion of personally identifiable utilization, it does not address the serial numbers attached with smart meter data measurements. Also, some important operating state data may be lost. Others \cite{varodayan2011smart} have proposed battery storage systems at the point of smart meter measurement to smooth, and thus destroy, utilization profiles. Such methods involve costly hardware, especially for the home customer. In none of these three investigations is the problem of data flood simultaneously addressed.

What is needed is an AMI architecture that allows the utility and customers to both benefit from AMI, yet anonymizes the data as necessary to protect individual privacy. The architecture presented serves to address both the data flood and privacy problems in a unique and mutually beneficial approach for the customer and the utility. 

\section{A New Architecture for AMI Deployment}
Existing AMI deployment topologies in AMI involve wireless, and occasionally wired, information networks from the smart meter to either a data concentrator, with no processing capabilities, or the utility directly. In all of these topologies, the utility receives all raw data and must archive and process this at a centralized point. Here, a new architecture is presented to address the data flood and privacy problems within a set of constraints that allow the architecture to preserve the functionality of existing AMI deployment. These constraints are as follows.
\begin{enumerate}[1.]
\item Customers must be ensured privacy of their real time utility service utilization.
\item Customers should have an opt-in path for individual analysis of their utility service.
\item Utilities must be reasonably relieved of the liability of maintaining private customer utilization data.
\item Utilities must be able to accurately perform load forecasting and estimation of operating state for the utility infrastructure.
\item Utilities must be able to reduce the AMI data set as much as practical to mitigate archiving and analysis effort.
\item Utilities must be able to individually bill for utility services, through automated meter reading, and control individual customer service connections.
\end{enumerate}
An example for a power utility is illustrated in Fig. 1. A subset of the AMI smart meters, which samples a subsection of the power grid, feeds local measurement data into the raw data buffer of an Aggregator via the AMI information network. The raw data buffer is a temporary buffer, which is periodically overwritten as smart meters broadcast their measurement data. This buffer provides the database from which to conduct an analysis of the local grid. Feature extraction is then performed on this local model analysis to build a report describing the operating state of that subsection of the power grid. The reports from all Aggregators become the basis for a whole grid operating state, as depicted in Fig. 2. This serves the dual purposes of reducing the data sent to the utility while filtering the private customer utilization profile. The Aggregator allows the architecture to conform to constraints 1, 3, 4, and 5.

\begin{figure}[htb]
\begin{center}
\includegraphics[]{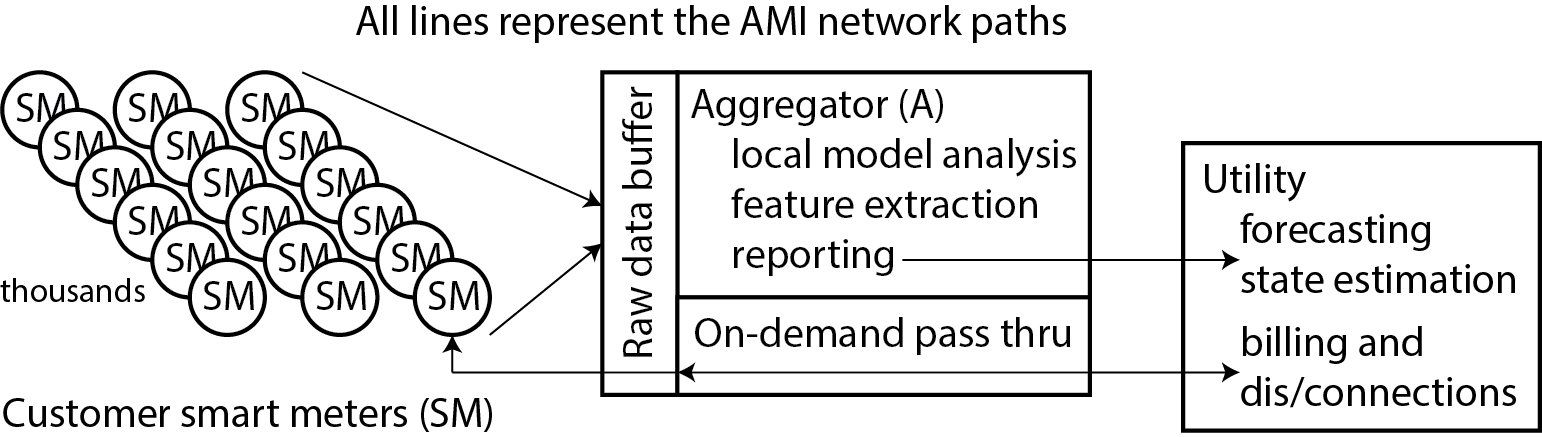}
\caption{Aggregation example for a power utility AMI.}
\end{center}
\end{figure}
\begin{figure}[htb]
\begin{center}
\includegraphics[]{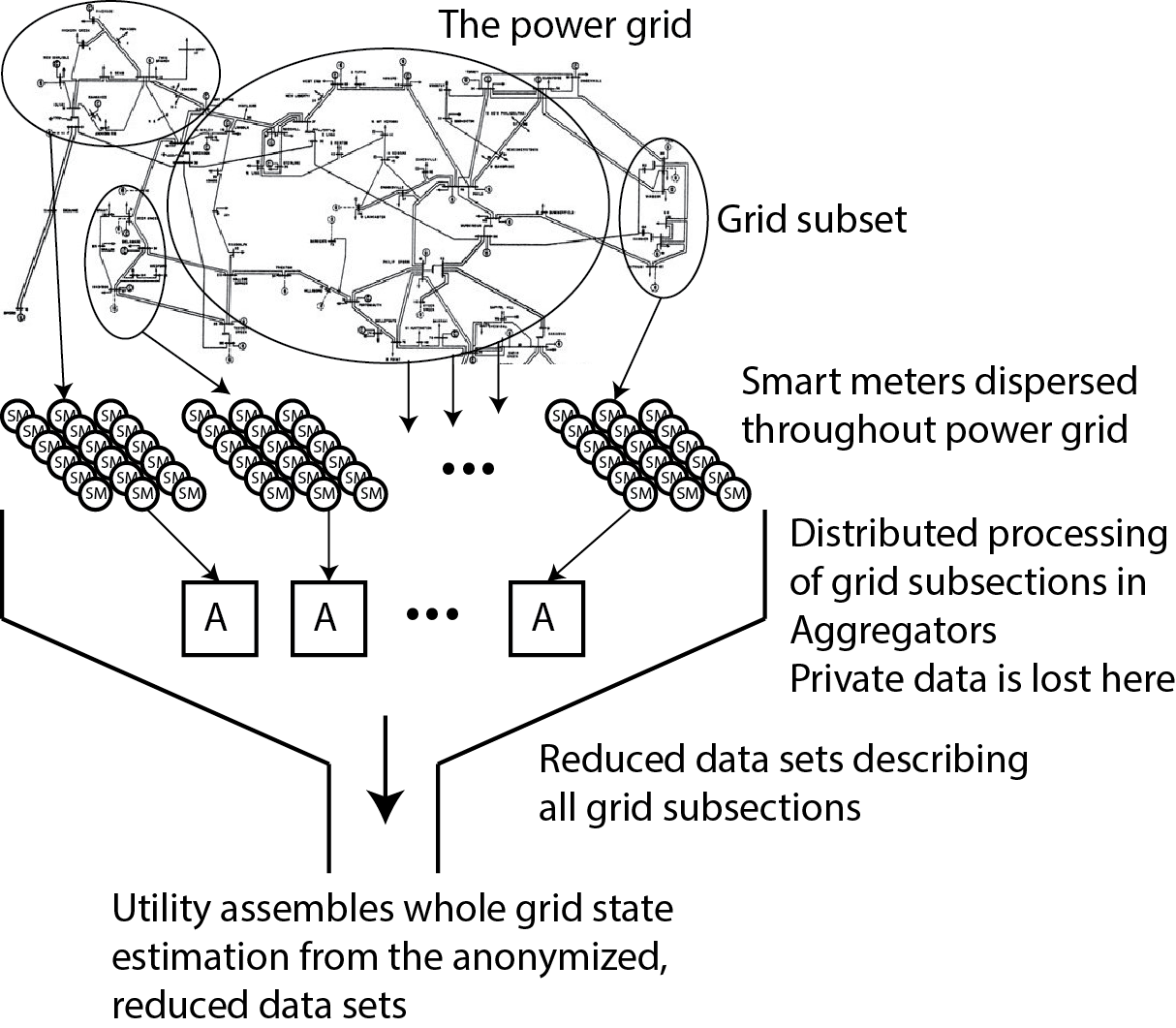}
\caption{Multiple Aggregators (A) sub-sectioning a utility grid, IEEE 118-bus test case.}
\end{center}
\end{figure}
Figure 1 also illustrates the approach for situations in which the utility needs to address an individual smart meter for billing and connection purposes through an on-demand pass thru. On demand, the utility may poll the raw data buffer to collect the cumulative power usage for an individual meter, thus achieving automated meter reading. This cumulated value contains no private utilization profile from which any additional personal insights can be inferred. Likewise, the utility may send on-demand connect/disconnect commands through this pass thru to suspend and restore local utility service as necessary. The on-demand pass thru conforms to constraints 1, 3, and 6.

In contrast to the data concentrators or direct links, which merely broadcast all raw data, the Aggregator in this architecture becomes a distributed processing node. Each Aggregator incorporates a model of that portion of the power grid monitored by its respective smart meter group. This modular approach allows a more granular analysis to be performed. Aggregators may also incorporate machine learning in order to self-tune their models for further fidelity. Features are extracted from this model based on a subset of points for operating state estimation, for example voltage, frequency, and current for a power grid. The features from this subset, rather than all raw data sets, are then sent to the utility, which assembles a whole grid operating state based on these modular estimations.

Constraint 2 poses a privacy exception, for if the customer wants to allow the utility to provide individual analysis of their service, such as detailed utility bills noting time-of-day utilization and inferred measures of appliance energy efficiency, the customer needs to provide such data to the utility. In this case, the customer may opt-in by pushing their data over their own Internet service to a server cloud from which the utility may access. Some commercial solutions \cite{vojdani2008smart,gungor2012smart} are beginning to surface that can provide reading and analysis of smart meter data to customers. Alternatively, the customer may keep such data and perform their own analysis locally without pushing to servers, should such commercial solutions become available. While not accessing the smart meter itself, TED\textsuperscript{\textregistered} The Energy Detective \cite{faruqui2010impact} is one such solution for local energy monitoring and analysis. Enabling such an opt-in service provides conformance to constraint 2. 

It should be noted that load forecasting follows from the ability to perform estimation of the total operating state, since the prediction of load is partially based on a history of operating states. This history is combined with more general knowledge, such as season, weather, time of day, day of week, or other societal factors, none of which introduce privacy issues nor add additional data requirements to the existing effort of load forecasting in this architecture versus the traditional case.
\section{Analysis of Architecture}
The architecture as specified has been logically demonstrated to meet the constraints introduced by definition. As such, an analysis of the architecture needs to demonstrate three metrics. In our previous power grid example, we validate the following.
\begin{enumerate}[1.]
\item Demonstrate that subsections of the power grid can be adequately represented by a subset of smart meter measurements. 
\item Demonstrate that this subset represents significant savings in data archiving resources. 
\item Demonstrate that any permanent data and all data collected by the utility contain no personally identifiable information of any individual customer.
\end{enumerate}
To demonstrate metric \#1, a residential subsection of the power grid representing a neighborhood with one hundred housing loads is simulated. The subsection consists of a single feeder at a service voltage of 240Vrms AC approximately 500m from the grid connection and 50m between each house, with service to each house branching from that feeder. Voltage is chosen to provide a simple illustration of data estimation, and because it is a close analog to pressure in gas and water networks. The expected voltage drop is illustrated in Fig. 3 driven by increasing load on the circuit chronologically by house number, and the total distance from the source. 
\begin{figure}[htb]
\begin{center}
\includegraphics[]{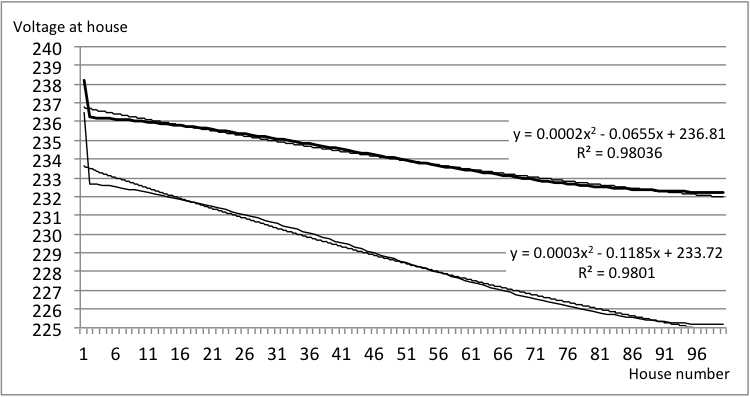}
\caption{Voltage drop throughout a 100-house neighborhood.}
\end{center}
\end{figure}
In Fig. 3, two cases are simulated. The lower line represents the case of all homes at 10kW of load and the subsequent voltage drop from 240V to approximately 225V. The upper line represents the case of all homes at a uniformly random load varying between 200W and 10kW. In both cases, a second order polynomial curve is fitted with $R^2 > 0.98$. The initial drop in voltage to the first house is due to the 500m run from the neighborhood feeder. Therefore, the voltage at any individual home may be closely estimated from the characterizing equation rather than relying on one hundred individual measurements of voltage. Knowledge of the equation representing voltage drop along this feeder does not provide insight into the load consumption at any single house. This mitigates both the data flood and privacy problems.

Metric \#2 follows from metric \#1, and could be inferred by observing that the data set reported to the utility, as in Fig. 2, is smaller than the original data set. This metric is also illustrated in Fig. 3 as a residential section of the power grid can be estimated through the use of a single equation with a few coefficients. Only the characterizing equation from the Aggregator is sent to the utility, thus dramatically reducing the data archiving and communications requirements. This is at least an order of magnitude reduction in data versus one hundred smart meter measurements. Figure 4 further illustrates how the grid is reduced to a network of subsections, with each subsection capable of being defined by an equivalent circuit model. This further mitigates the data flood problem.
\begin{figure}[htb]
\begin{center}
\includegraphics[]{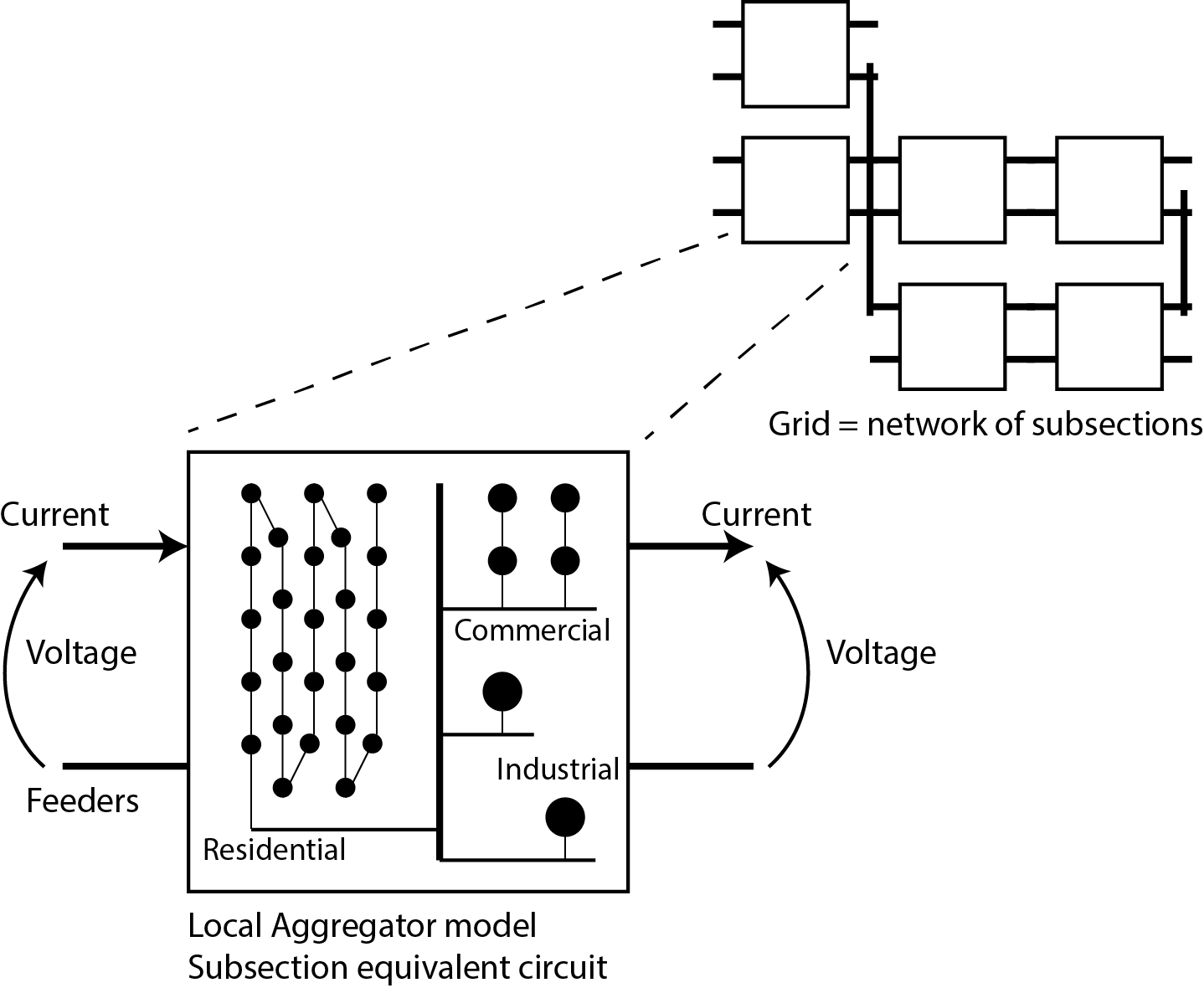}
\caption{Reduction factors for operating state estimation in a tiny electric utility network.}
\end{center}
\end{figure}
Metric \#3 is also demonstrated by observation of Figs. 1 and 2 in that all individually identifiable information is associated with the smart meter raw data. The data set of each smart meter retains the serial number, and thus personal id, of that customer. Since this information is only saved in a temporary buffer in the Aggregator, it will be automatically overwritten before there is sufficient history to determine any significant behavior patterns from an individual smart meter data set. In the event an attacker compromises an Aggregator, the data residing in the buffer will often be too short in duration for meaningful long-term prediction of customer behavior. Aggregated data representing the local features for the operating state of that AggregatorÕs section of the power grid will consist of an aggregated approximation of the model, rather than any single smart meter measurement. In Eq. 1, the operating state matrix, $\mathbb{O}$, is determined by a function of each smart meter measurements for that subsection of the power grid. Vector, $\vec{m}_n$, represents the measurement of the $n^{th}$ smart meter. These vectors are aggregated into the matrix, $\mathbb{X}$, but only the operating state is reported to the utility as in Eq. 2. In order to extract individual measurement data, as in Eq. 3, an inverse function would be necessary, which does not exist. Further, since the operating state contains an aggregation of individual smart meter measurements through a nonlinear function, there is not sufficient dimension from which to extract these individual measurements. Thus in Eq. 4, it is not possible to recreate $\mathbb{X}$ from $\mathbb{O}$, even if an inverse function was discovered. The procedure is as follows.\\
The operating state is determinedÉ
\begin{equation}
\mathbb{O} = \mathbb{F} \lbrace \vec{m}_1 , \vec{m}_2 , ... , \vec{m}_N \rbrace
\end{equation}
and only the operating state is reported to utility.\\
\begin{equation}
\mathbb{F} \lbrace \mathbb{X} \rbrace = \mathbb{O}
\end{equation}
The inverse of the operating state is not known from which to recover $\mathbb{X}$É\\
\begin{equation}
\mathbb{F}^{-1} \lbrace \mathbb{O} \rbrace = \mathbb{X}
\end{equation}
and $\mathbb{X}$ cannot be broken down into individual measurements due to aggregation.\\
\begin{equation}
\mathbb{X} \nRightarrow \lbrace \vec{m}_1 , \vec{m}_2 , ... , \vec{m}_N \rbrace
\end{equation}

The individual smart meter measurement vectors are never reported to the utility and are locally overwritten in a relatively short time period, e.g., in hours. This further mitigates the privacy problem.

\section{Conclusions}
The demonstration of the three metrics is applied to the power grid AMI example, yet all utility service AMI would follow in a similar manner. For water networks, pipe pressures and flows serve as the measurement data, similarly for natural gas networks. It is demonstrated that the data flood to the utility can be mitigated without a loss of fidelity in the estimation of operating state or in load forecasting. Data archiving requirements would be reduced by a substantial factor, especially in dense, residential deployments were measurement redundancy is strong. The distributed field processing of the grid model in Aggregators may result in reduced computing requirements overall, but at least certainly at the utility level. Communications issues would be significantly reduced moving to this architecture versus the existing case of many thousands of smart meters pushing data to a single centralized utility server. 

Likewise, data privacy can be maintained to the satisfaction of the customer. The maintenance of data privacy not only reduces liability to the utility, but also facilitates data retrieval and analysis by utility scientists and engineers. In many cases, these analysts must go through special procedures to gain power grid data because it is currently attached to personally identifiable information. Eliminating the cumbersome procedure of obtaining proper authorizations to view personally identifiable information would be a significant benefit.

Given that most AMI networks require some form of distributed data concentrators, or at least data relay points, the substitution of Aggregators as illustrated in Fig. 1 should have a relatively neutral hardware cost impact. Software cost may increase on the field side for grid subsection models, but it is expected that this cost would be balanced by reduced data archiving and processing requirements in the utility's central office \cite{joskow2012creating}.

\section{Future Work}
A new cyber range capability is being developed, which will allow the emulation of AMI deployments at the scale of many thousands. This work will involve the development of virtual machines that emulate smart meter hardware and AMI network infrastructure. Such deployments have not been simulated in the past due to the large resources needed for widespread AMI networks beyond about $10^3$ in scale. In many studies, an authentic representation of AMI is not possible without the simulation being done at a representative scale, i.e., AMI simulations of tens or even hundreds of smart meters does not provide insights into how such deployments function at scales of $10^4$ to $10^6$.

\bibliographystyle{unsrt}
\bibliography{AMI-aggregation}

\end{document}